\documentclass[a4paper,11pt]{article}
\usepackage{comment}
\usepackage{ulem}
\pdfoutput=1 

\usepackage{jcappub} 

\usepackage[T1]{fontenc} 
\usepackage{orcidlink}

\title{\boldmath Thermodynamics of an universe with Decaying Cold Dark Matter}

\author{Javier Ju\'arez-Jim\'enez\orcidlink{0009-0004-8314-7776}$^1$, Ana A. Avilez-López\orcidlink{0000-0003-2223-4716}$^{1,2}$ and Miguel Cruz\orcidlink{0000-0003-3826-1321}$^3$}
\emailAdd{javier.juarezji@alumno.buap.mx}
\emailAdd{ana.avilezlopez@correo.buap.mx}
\emailAdd{miguelcruz02@uv.mx}

\affiliation{$^1$Facultad de Ciencias F\'{\i}sico Matem\'aticas, Benem\'erita Universidad Aut\'onoma de Puebla, Apdo. Postal 1152, Puebla, Pue., M\'exico, \\
$^2$Centro Internacional de F\'isica Fundamental, Benem\'erita Universidad Aut\'onoma de Puebla, Apdo. Postal 1152, Puebla, Pue., M\'exico,\\
$^3$Facultad de F\'{\i}sica, Universidad Veracruzana 91097, Xalapa, Veracruz, M\'exico.}

\abstract{In this work we focus on the thermodynamics consistency of a new set of solutions emerging from a cosmology in which dark matter is able to decay into relativistic particles within the dark sector. It is important to stress that the lifetime of dark matter is larger than the age of the universe in order to be consistent with observations. Given that the corresponding decay rate is small, this one can be used as a perturbative parameter and it is possible to construct analytic solutions from a perturbative analysis for the densities of the species and the scale factor. The decay of dark matter is an irreversible process since it occurs out of chemical equilibrium and therefore the entropy per comoving volume increases considerably, as a consequence the temperature does not scale as $a^{-1}$ in contrast to an adiabatic expansion. We take into account two scenarios: a) The case in which both species making up the fluid end up in thermal equilibrium and therefore their temperature is the same. b) A second instance in which the species do not reach thermal equilibrium and therefore they have different temperatures. We verify that the second law of thermodynamics is satisfied in any case.}

\begin{document}

\maketitle

\section{Introduction}
\label{sec:intro}

It is well known that the Standard Model of Cosmology $\Lambda$CDM is able to predict the accelerated expansion of the universe and provides a good description of the process of structure formation at large scales at late times. Furthermore, it provides a stunningly precise prediction of the temperature and polarization anisotropies of the cosmic background radiation (CMB) decoupled from the primordial plasma at early times \textbf{\citep{Planck2018,Verde2019}}. Within this model, cold dark matter makes up around a fourth of the total matter and energy content in the universe, which is prescribed as a non-collisional perfect fluid that is affected by other species of particles only by means of gravity. 

Observations of rotation curves of spiral galaxies \textbf{\citep{VeraRubin1970}}, the dynamics of clusters, the distribution of matter at cosmological scales, and the anisotropies of the CMB  \textbf{\citep{Planck2018}} suggest the presence of dark matter at different scales and ages of the universe. Moreover, observations of the distribution of matter in the bullet cluster \textbf{\citep{Douglas2004, Marusa2008}}, measurements of the anisotropies of the CMB, and the study of the structure at large and intermediate scales \textbf{\citep{Blumenthal1984}} indicate that this component definitively interacts with itself and other species gravitationally, while other forms of interaction are considered absent or negligible. For those reasons, within the standard $\Lambda$CDM paradigm, dark matter is modeled as a non-collisional perfect fluid \textbf{\citep{Moore2000}}.

However, some aspects give rise to some tensions within the $\Lambda$CDM model and observations. A remarkable example is that there is a $5\sigma$ discrepancy between estimates of the Hubble parameter $H_0$ inferred from SNaI supernovae data and $\Lambda$CDM inferences from early universe data such as measurements of the sound horizon at recombination from the temperature spectra of the CMB anisotropies  \textbf{\citep{Verde2019, Planck2018}}. Besides, within the standard paradigm, dark matter is presumably described as a pressureless perfect fluid dubbed as dust. Although this prescription provides a sufficiently good description of the process of structure formation and the nowadays distribution of matter at linear cosmological scales, there are still open questions regarding what happens at the non-linear regime at small scales \textbf{\citep{Ostriker2003,BullockKolchin2017,Schneider2015}}. Among the possible issues, the cusp-core, the missing satellite, and the too-big-to-fail problems can be cited \textbf{\citep{Jiang2015,Moore1999,Moore1998,Klypin1999}}. Various of these problems are overcome with the inclusion of baryonic feedback; however, there remain open questions regarding systems with low surface brightness in which baryons are not too abundant. The previous fact suggests that possibly dark matter might behave differently than a non-collisional perfect fluid, at least at small scales. In this regard, it is worth mentioning that these possible discrepancies between theory and observations are absent in a host of  alternative scenarios for the description of dark matter. To mention a representative set, fuzzy dark matter and generalized dark matter \textbf{\cite{Hui2016,Hu_1998}} feature a cutoff in the matter perturbations power spectrum at small scales, which prevents the proliferation of very small structures and mechanisms to flatten the centers of galaxy halos, regardless of baryonic feedback. Among the scenarios in which potential inconsistencies of cold dark matter are overcome, we highlight those in which dark matter is described as a fluid with more general properties than dust, see for instance \textbf{\citep{doi:10.1142/S0218271804006346}}. The simplest extension would be that dark matter is a fluid with pressure arisen from internal collisional processes within the dark sector. In this fashion, we can mention some scenarios in which properties of dark matter as a fluid are modified due to different types of interactions, for example, self-interacting dark matter and decaying dark matter in which a thermal transfer of energy and momentum between dark matter and dark radiation occurs due to an out of equilibrium decaying process. Another widely studied model is that of fuzzy dark matter, in which a quantum pressure contribution, due to the wave behavior of dark matter, produces modifications to the standard dust-like fluid behavior \textbf{\citep{Tulin2017,Audren:2014bca,Hui2016}}. In this work, we are interested in studying the thermodynamics of the decay of dark matter in this last kind of model. Within this scenario, some cosmological tensions such as the $H_0$ tension and the $\sigma_8$ tension are alleviated up to some extent. Another interesting scenario takes into account the possibility that dark matter is made of multiple species which may interact with each other but not with the particles of the standard model. The parametrized phenomenological framework called Generalized Dark Matter (GDM) originally proposed by Hu \textbf{\citep{Hu_1998}} and widely studied by Kopp et al. \textbf{\citep{Kopp2016}} has the advantage of enclosing a large family of models in which dark matter is described as a fluid with adiabatic pressure (either at the background level and the perturbations) and viscosity. Recently, predictions of counts of Hickson's galaxy groups in the local universe have been obtained within this framework and turn out to be consistent with observations in contrast to those for the standard cold dark matter scenario \textbf{\citep{Lopez-Sanchez2022}}.  

This article is structured in the following way. In the section \ref{sec:ddm} we present the cosmological model considered, which corresponds to a simple extension of the standard cosmological model in which cold dark matter is allowed to interchange energy with a relativistic specie belonging to the dark sector dubbed as dark radiation. Since the decaying width of dark matter should be sufficiently small in order to give rise to dark matter lifetimes of cosmic scales, the continuity equations corresponding to the dark species involved in the decaying process can be solved perturbatively. In section \ref{sec:analytical} we construct analytic solutions for the energy densities of decaying dark matter, dark radiation, and the scale factor. In section \ref{sec:thermo}, we study the thermodynamics of this kind of universe, given that during the expansion history the decaying process of dark matter produces a significant amount of entropy per comoving volume. The temperatures of the dark sector components are allowed to evolve, having a crucial distinction from the standard cold dark matter model. Section \ref{sec:final} is devoted to the final comments of our work.

\section{Decaying dark matter model}
\label{sec:ddm}

In this section we consider a model which generalizes the assumptions considered for the dark matter sector in the $\Lambda$CDM description; in this case the stability condition for the cold dark matter sector is relaxed. We allow cold dark matter (DCDM) species to decay into relativistic particles belonging to the dark sector dubbed as {\it dark radiation} (DR). For instance, this kind of decay for dark matter was used to provide an explanation for the observed excess of gamma rays in some galactic centers \textbf{\citep{dmdm1}}. For our analysis we consider for DCDM a mean lifetime of the order of cosmic scales in order to be consistent with astrophysical observations, see for instance Ref. \textbf{\citep{diff2}}, where it was shown that late transitions from DCDM to DR do not alter severely the temperature of the anisotropy spectrum.  

\subsection{Continuity equations for the energy densities of dark species}

Concerning the general properties of established particles in nature, absolute stability represents an exceptional case within the current standard model of particle physics; such stability necessitates the presence of exact symmetries. Consequently, this characteristic may be attributed to potential dark matter particle candidates without contradicting observed behaviors among the fundamental constituents of matter.

In contrast to the standard scenario where all species of particles are thermally decoupled, within the model considered here, the continuity equation governing the evolution of DCDM energy density turns out to have a positive (negative) collision term counting for the loss (gain) of energy due to the direct (inverse) decay. Besides, the continuity equation for DR holds a positive (negative) collision term given that this specie gains (loses) energy while particles are produced (destroyed). In this sense, the continuity equation for each specie is given as
\begin{align}
    & \dot{\rho}_{\mathrm{dcdm}}+3H\rho_{\mathrm{dcdm}}=-\epsilon_{\mathrm{dcdm}} \rho_{\mathrm{dcdm}},
    \label{density_1}\\
    & \dot{\rho}_{\mathrm{DR}}+4H\rho_{\mathrm{DR}}=\epsilon_{\mathrm{dcdm}} \rho_{\mathrm{dcdm}},
    \label{density_2}
\end{align}
where $\rho_{\mathrm{dcdm}}$ is the energy density of DCDM, $\rho_{\mathrm{dr}}$ is the energy density of DR, $\epsilon_{\mathrm{dcdm}}$ is the DCDM decay rate (and $\epsilon^{-1}_{\mathrm{dcdm}}$ its lifetime) and $H\equiv \dot{a}/a$ is the Hubble parameter. From now on, dots indicate derivatives with respect to proper time. As can be seen from the r.h.s. of the set of equations given above, the collision terms of each species are proportional to the energy density of dark matter; this dependence arises from the assumption that a single particle of dark matter decays into a combination of radiation and massive particles. However, we restrict ourselves to the case of relativistic decay products. In section \ref{sec:analytical} it will be shown that it is possible to solve analytically the former system of equations in the limit where the mean DCDM lifetime is greater than the age of the universe, or equivalently, when the decay rate $\epsilon_{\mathrm{dcdm}} \simeq 0$.

\subsection{Conservation equations for the number densities of dark species}

On the other hand, we also have a system of equations that describes the number densities of the dark species: 
\begin{align}
    & \dot{n}_{\mathrm{dcdm}} + 3Hn_{\mathrm{dcdm}}= \Gamma_{\mathrm{dcdm}}n_{\mathrm{dcdm}},\\
    & \dot{n}_{\mathrm{DR}} + 3Hn_{\mathrm{DR}}= \Gamma_{\mathrm{DR}}n_{\mathrm{DR}},
\end{align}
where the quantity $\Gamma_{\mathrm{dcdm}}$ accounts for the rate of change of the number of particles of DCDM (note that $\Gamma_{\mathrm{dcdm}}<0$), while $\Gamma_{\mathrm{DR}}$ represents the corresponding production rate of DR particles, and it must be a positive number. From equation \ref{density_1} we can see that $\Gamma_{\mathrm{dcdm}} = -\epsilon$, indicating that cold dark matter is decaying, as desired. In contrast, in order to obtain an expression for $\Gamma_{\mathrm{DR}}$ we assume that for every DCDM decay, two particles of DR are produced, in analogy with previous works, for example, where the dark matter candidate is the majoron, and the leading decay channel is two relativistic neutrinos \textbf{\citep{valle1, valle2, Audren:2014bca}}. With this, $\Gamma_{\mathrm{DR}}$ is given by
\begin{equation}
   \Gamma_{\mathrm{DR}} = \frac{1}{2} \frac{n_{\mathrm{dcdm}}\epsilon}{n_{\mathrm{DR}}}.
\end{equation}
It is important to mention that the scenario considered in this work differs from those reported in Refs. \textbf{\citep{diff1, diff2, diff3}}. Therefore, the analytical solutions discussed below represent a new set of solutions for the cosmic components in this kind of cosmological scenarios.

\section{Analytical background solutions for the energy densities of decaying dark matter and dark radiation}
\label{sec:analytical}
In this section, we implement a perturbative method to solve the system of equations \ref{density_1} and \ref{density_2} in order to obtain analytical solutions for the energy densities of DCDM and DR at the background level. 

\subsection{Perturbative construction of the background solutions}
\begin{figure}[h!]
        \centering
        \includegraphics[scale=0.52]{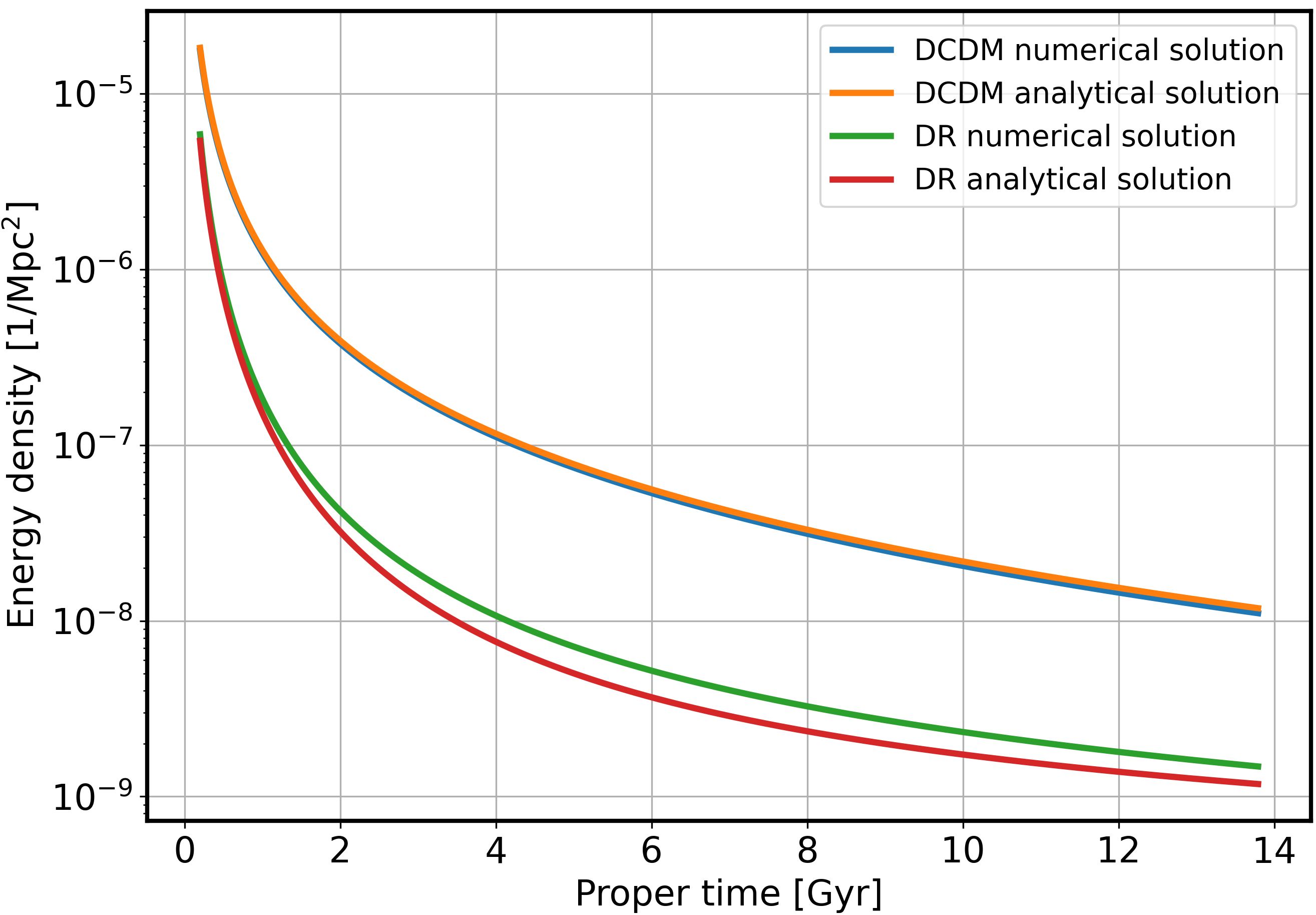}
        
        \caption{Evolution in terms of proper time for the DCDM energy density and DR energy density with  $\eta_i=2\sqrt{\gamma/\hat{\rho}_{\mathrm{dcdm}_0}}, \eta_f =\hat{\rho}_{\mathrm{dcdm}_0}^{1/2}(\sqrt{\gamma+1}-\sqrt{\gamma}),\gamma=0.03$. }
        \label{densities}
    \end{figure}

We restrict ourselves to the limit where the average lifetime of DCDM is greater than the age of the universe, allowing us to propose solutions for $\rho_{\mathrm{dcdm}}$, $\rho_{\mathrm{DR}}$ and the scale factor, as follows, since in this case $\epsilon_{\mathrm{dcdm}}\ll 1$:
\begin{align}
    & \rho_{\mathrm{dcdm} }=\bar\rho_{\mathrm{dcdm}} + \delta\rho_{\mathrm{dcdm}},\\
    & \rho_{\mathrm{DR}} = \bar\rho_{\mathrm{DR}}+\delta\rho_{\mathrm{DR}},\\
    & a = \bar a + \delta a,
\end{align}
where the bar stands for the well-known stable solutions; in this sense, the decay of DCDM particles induces changes in the cosmological quantities, which we will characterize as a small perturbation denoted as $\delta$. Inserting the quantities given above into the system of equations \ref{density_1} and \ref{density_2} and the Friedmann equation 
\begin{equation}
   \left(\frac{\dot{a}}{a} \right)^2  = \frac{8\pi G}{3}\left( \rho_{\mathrm{dcdm}}+ \rho_{\mathrm{DR}}\right) \label{ec_friedmann},
\end{equation}
one gets the following expressions for the unperturbed quantities
\begin{align}
    & \dot{\bar\rho}_{\mathrm{dcdm}}+3\left(\frac{\dot{\bar a}}{\bar a}\right) \bar\rho_{\mathrm{dcdm}} = 0,\\
    & \dot{\bar\rho}_{\mathrm{DR}}+4\left(\frac{\dot{\bar a}}{\bar a}\right) \bar{\rho}_{\mathrm{DR}} = 0,\\
    & \bar{H}^2 \equiv  \bigg(\frac{\dot{\bar a}}{\bar a}\bigg)^2 = \frac{8\pi G}{3} (\bar \rho_{\mathrm{dcdm}}+\bar\rho_{\mathrm{DR}} ), \label{friedmann_zero}
\end{align}
whose solutions are already known. At first order in the perturbations, we can write the Friedmann constraint as
   \begin{equation}
    2\bigg(\frac{\dot{\bar a}}{\bar a}\bigg)^2 \bigg(\frac{\delta \dot{a}}{\dot{\bar a}} -\frac{\delta a}{\bar a}\bigg) =  \frac{8\pi G}{3} (\delta\rho_{\mathrm{dcdm}} +\delta\rho_{\mathrm{DR}} ),
\end{equation}
and for the energy densities of the dark sector, we have
\begin{align}
& \delta\dot{\rho}_{\mathrm{dcdm}} + 3\bar{H} \delta\rho_{\mathrm{dcdm}}\bigg(1+\frac{1}{2}\bar\Omega_{\mathrm{dcdm}}\bigg) +\frac{3}{2}\bar{H}\bar\Omega_{\mathrm{dcdm}}\delta\rho_{\mathrm{DR}} = -\bar\rho_{\mathrm{dcdm}}\epsilon_{\mathrm{dcdm}}\label{sist_prev1},\\
& \delta\dot{\rho}_{\mathrm{DR}} + 4\bar{H} \delta\rho_{\mathrm{DR}} \bigg(1+\frac{1}{2}\bar\Omega_{\mathrm{DR}}\bigg) +2\bar{H}\bar\Omega_{\mathrm{DR}}\delta\rho_{\mathrm{dcdm}} = \bar\rho_{\mathrm{dcdm}}\epsilon_{\mathrm{dcdm}} \label{sist_prev2},
\end{align}
where $\bar\Omega_{\mathrm{dcdm}} = \frac{\bar\rho_{\mathrm{dcdm}}}{\rho_c} = \bar\rho_{\mathrm{dcdm}_0} \bar a^{-3} \frac{8\pi G}{3\bar{H}^2}$,  $ \bar \Omega_{\mathrm{DR}} = \frac{\bar\rho_{\mathrm{DR}}}{\rho_c} = \bar\rho_{\mathrm{DR}_0} \bar a^{-4} \frac{8\pi G}{3\bar{H}^2}$ are the relative densities of each dark species,  $\rho_c$ is the critical density and quantities with a subscript '0' denote their present-day values. At this point, it is convenient to introduce a new time variable, called \textit{conformal time}, usually denoted by $\eta$ and given by the expression
\begin{equation}
    \eta = \int \frac{dt}{\bar a(t)}.
\end{equation}
By considering the change of variable given above, we solve for the scale factor using the equation \ref{friedmann_zero}, yielding 
\begin{equation}
   \bar a(\eta)= \hat{\rho}_{\mathrm{dcdm}_0}\eta^2 + 2\sqrt{\hat{\rho}_{\mathrm{dcdm}_0}\gamma}\eta, \label{scale_factor}
\end{equation}
where $\hat{\rho}_{\mathrm{dcdm}_0}\equiv \frac{2\pi G}{3}\bar\rho_{\mathrm{dcdm}_0}$ and $\gamma   \equiv \frac{\bar\rho_{\mathrm{DR}_0}}{\bar\rho_{\mathrm{dcdm}_0}}$. Therefore, $\gamma$ is the fraction of DR with respect to DCDM at present time, notice that this value must be close to zero in order to be consistent with the standard model. On the other hand, $\gamma$ must be related to the lifetime of DCDM, indeed we can obtain a relation between these quantities by the long-lived DCDM hypothesis, assuming that the energy density of DR does not change significantly, then equation \ref{density_2} evaluated at present time gives us the condition
\begin{equation}
    \epsilon_{\mathrm{dcdm}} = 4H_0\gamma,
\end{equation}
where $H_0$ is the Hubble constant. 
At first order in the perturbations, the system \ref{sist_prev1} and \ref{sist_prev2} can be decoupled, yielding the following solutions for the energy densities 
\begin{align}
    & \rho_{\mathrm{dcdm}}(\eta)=\frac{4\hat{\rho}_{\mathrm{dcdm}_0}}{\left(\hat{\rho}_{\mathrm{dcdm}_0}\eta^2+2\sqrt{\hat{\rho}_{\mathrm{dcdm}_0}\gamma}\right)^3}+ 4\epsilon_{\mathrm{dcdm}} \hat{\rho}_{\mathrm{dcdm}_0}^{-1} \left[-\frac{4}{15}\eta^{-3}+\frac{4}{5}\sqrt{\frac{\gamma}{\hat{\rho}_{\mathrm{dcdm}_0}}}\eta^{-4}\right]\label{analitic_1},\\
    & \rho_{\mathrm{DR}}(\eta)=\frac{4\hat{\rho}_{\mathrm{dcdm}_0}\gamma}{\left(\hat{\rho}_{\mathrm{dcdm}_0}\eta^2+2\sqrt{\hat{\rho}_{\mathrm{dcdm}_0}\gamma}\right)^4}+ 4\epsilon_{\mathrm{dcdm}} \hat{\rho}_{\mathrm{dcdm}_0}^{-1} \left[\frac{1}{5}\eta^{-3}-\frac{3}{5}\sqrt{\frac{\gamma}{\hat{\rho}_{\mathrm{dcdm}_0}}}\eta^{-4}\right]\label{analitic_2}.
\end{align}
The former expansions are valid only when $\eta > 2\sqrt{\gamma/\hat{\rho}_{\mathrm{dcdm}_0}}$. 
Numerical solutions for the system \ref{density_1} and \ref{density_2} were also obtained and compared to the analytical solutions \ref{analitic_1} and \ref{analitic_2} in order to validate our approximation. As can be seen in Figure \ref{densities}, both results are in good agreement. In the subsequent sections, we implement our analytical solutions to characterize the thermodynamical processes governing this cosmological model. Viable cosmological models must demonstrate rigorous physical consistency, as an example, by adhering to fundamental thermodynamic principles, which serve as essential criteria for evaluating the theoretical soundness and observational compatibility of any proposed framework, see for instance \textbf{\citep{us2}}.

\section{Thermodynamics of the decaying process}
\label{sec:thermo}

Dark matter decay occurs beyond thermodynamic equilibrium through interactions between dark sector species. Due to the existence of such interactions, this non-equilibrium process is effectively modeled using dissipative fluid formalism, as in the GDM scenario \textbf{\citep{Hu_1998}}, providing a mathematical framework to derive coupled temperature evolution equations for the interacting components. This approach incorporates entropy production rates while connecting microscopic dark sector physics to observable cosmological phenomena. 

\subsection{Out of equilibrium decaying process}
The energy exchange between two cosmological fluids with non-conserved particle numbers leads to entropy production. As entropy is a state function, this process determines the evolution equation for the fluids' temperature as follows \textbf{\citep{8157526}}
\begin{equation}
        \dot{T}_A = -T_A\left(3H-\Gamma_A\right)\frac{\partial p_A/\partial T_A}{\partial \rho_A/\partial T_A} + \frac{\epsilon_A \rho_A -\Gamma_A(\rho_A + p_A)}{\partial \rho_A/\partial T_A}.
\end{equation}
Here it is convenient to associate the effective equation of state parameter related to the fluid $\omega_{\mathrm{eff}}=\frac{\partial p_A/\partial T_A}{\partial \rho_A/\partial T_A} = P_A / \rho_A$ in such a way that we can write 
\begin{equation}
        \dot{T}_A = -T_A\left(3H-\Gamma_A\right)\omega_{\mathrm{eff}} + \frac{\epsilon_A \rho_A -\Gamma_A\rho_A(1 + \omega_{\mathrm{eff}})}{\partial \rho_A/\partial T_A} \label{tempe}
\end{equation}
where $\epsilon_A = \epsilon$ for DR and $\epsilon_A= -\epsilon$ for DCDM, as stated in equations \ref{density_1} and \ref{density_2}. As can be seen from the expression given above, the cases $\epsilon_{A}, \Gamma_{A} = 0$ recover the expression obtained in the context of equilibrium thermodynamics \textbf{\citep{maartens}}
\begin{equation}
    \frac{\dot{T}_{A}}{T_{A}} = -3H\left(\frac{\partial p_{A}}{\partial \rho_{A}} \right)_{n_{A}},
\end{equation}
in which $\rho_{A}$ and $n_{A}$ are conserved. 

\subsubsection{The equation of state parameters of different (mixed) species}

\begin{figure}[htbp!]
        \centering
        \includegraphics[scale=0.5]{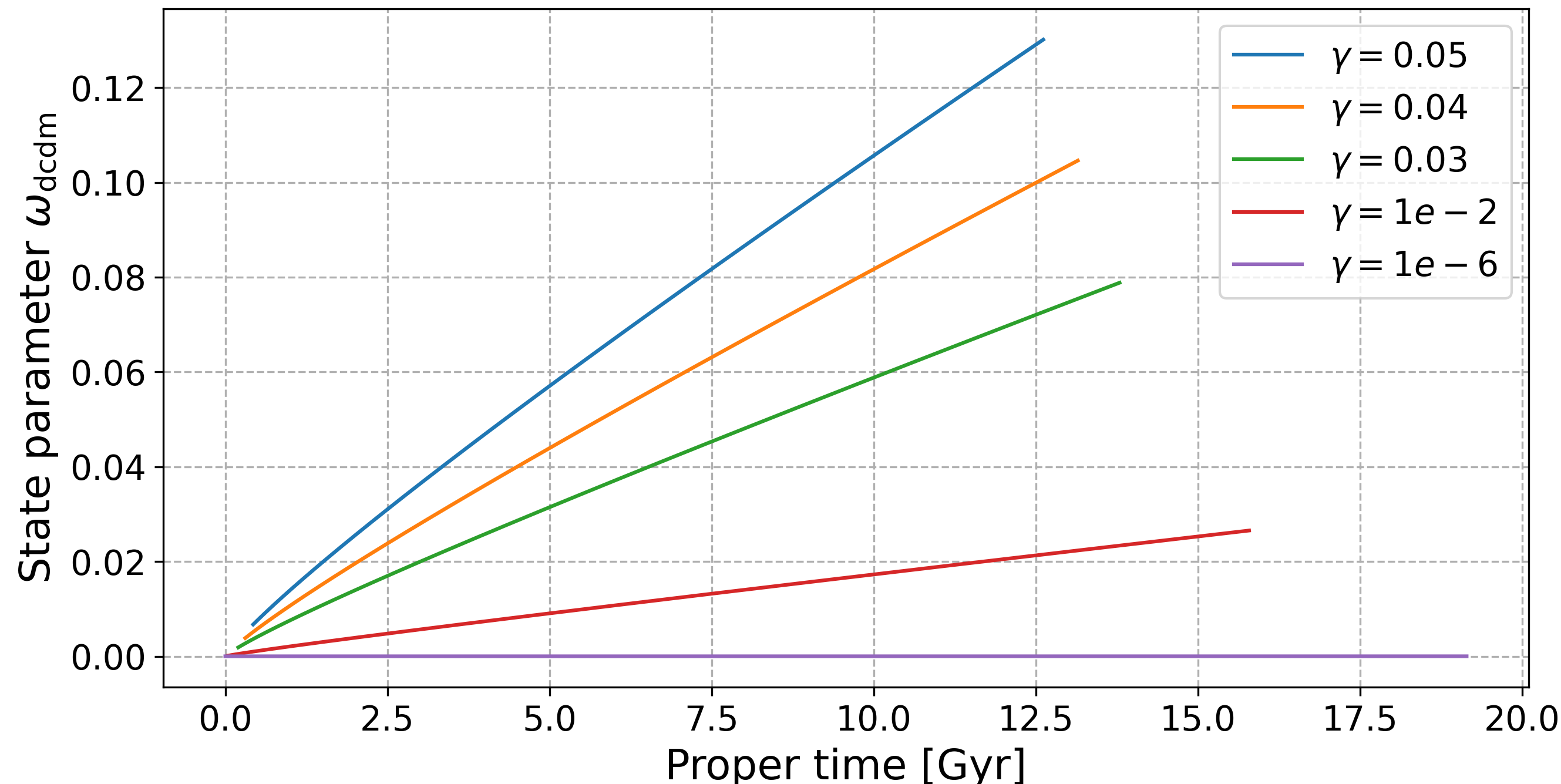}
        \includegraphics[scale=0.5]{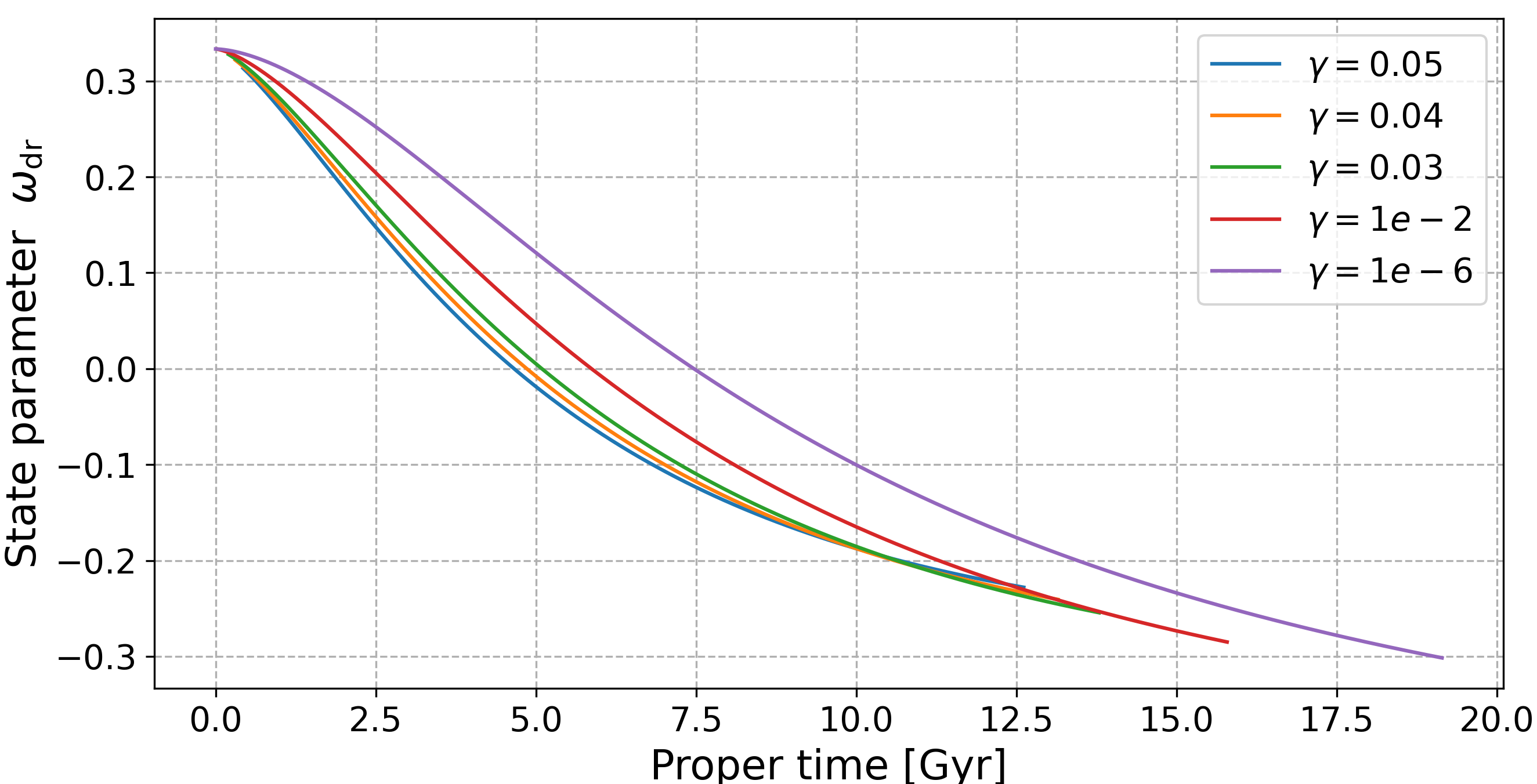}
        \includegraphics[scale=0.5]{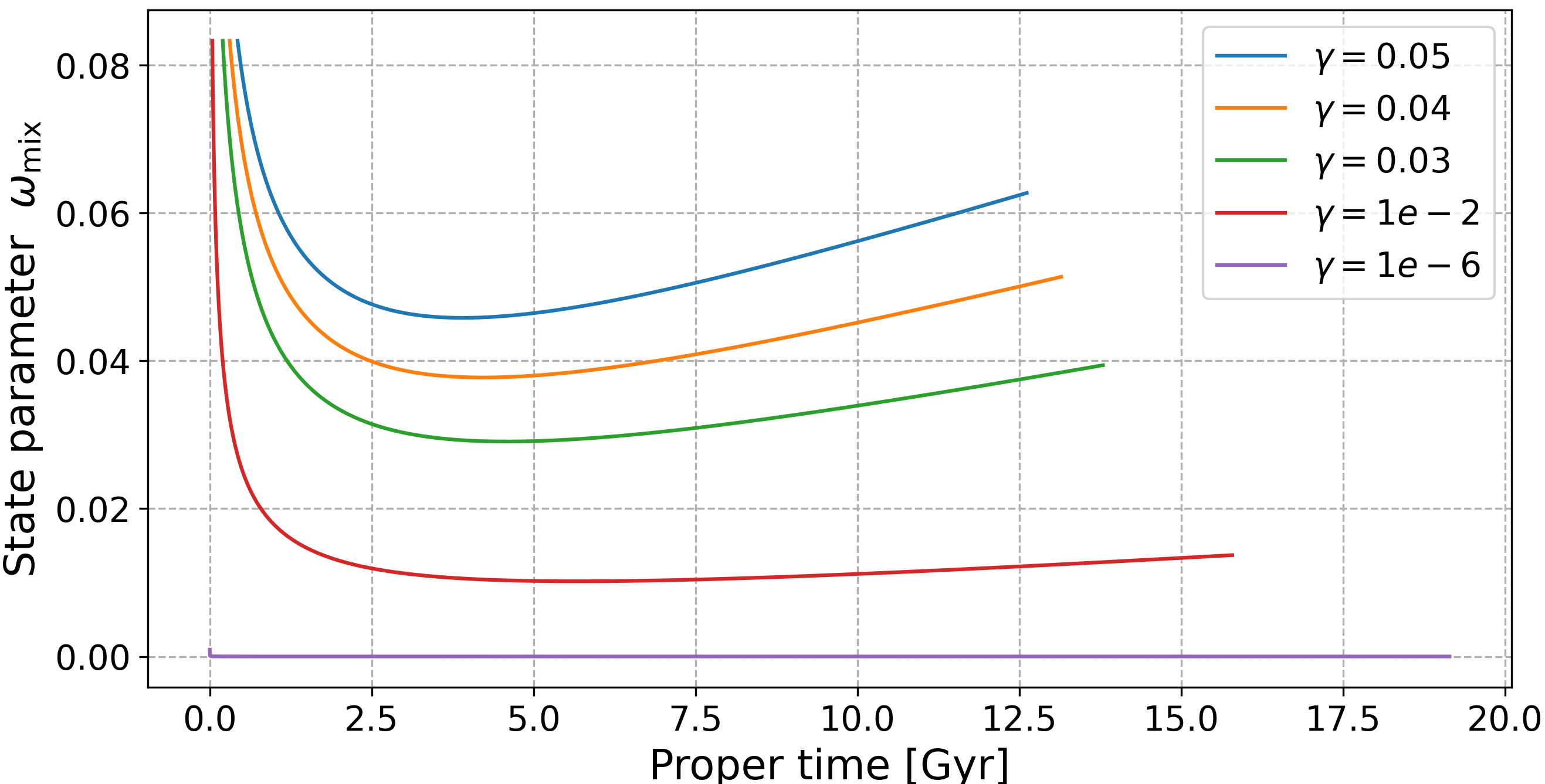}
        \caption{Behavior of the dynamical effective equation of state parameter.}
        \label{fig:omegas}
\end{figure}

Before proceeding to the temperatures analysis, we will comment about the corresponding equation of state parameters for the species involved. As usually done, we can define an effective equation of state parameter for the cosmic fluids by writing each equation of the system \ref{density_1} and \ref{density_2} in the standard form 
\begin{equation}
	\dot{\rho} + 3 H \rho(1+\omega_{\mathrm{eff}}) = 0 \label{ecuacion_continuidad},
\end{equation}
therefore the effective equation of state parameter of DCDM is given as
\begin{equation}
\omega_{\mathrm{dcdm}} = \frac{\epsilon_{\mathrm{dcdm}} }{3H}, \label{state_dcdm}
\end{equation}
and for DR we have,
\begin{equation}
\omega_{\mathrm{DR}} = \frac{1}{3} - \frac{\epsilon_{\mathrm{dcdm}}  \rho_{\mathrm{dcdm}}}{3H\rho_{\mathrm{DR}}}. \label{state_dr}
\end{equation}
It is worthy to mention that in this scenario the fluids have variable equation of state parameters, as can be seen in Figure \ref{fig:omegas}, instead of constant values as in the standard case: $\omega_{\mathrm{cdm}}=0$ and $\omega_{\mathrm{r}}=1/3$. Notice that for $\omega_{\mathrm{DR}}$ there exists a transition from positive to negative values at some stage of the cosmic evolution. For $\epsilon_{\mathrm{dcdm}} = 0$ (no decay) the usual values are recovered from the expressions given above. The dynamical behavior of the equation of state parameter per component has an impact on the cosmological evolution, as can be seen in the acceleration equation  
\begin{equation}
    \frac{\ddot{a}}{a}  = -\frac{4\pi G}{3} (\rho_{\mathrm{dcdm}}(1+3 \omega_{\mathrm{dcdm}})+\rho_{\mathrm{DR}}(1+3 \omega_{\mathrm{DR}})).
\end{equation}
Additionally, if we consider both species as a single fluid, we can define the equation of state parameter for such mix from the use of equations \ref{density_1} and \ref{density_2}, therefore this mix represents the case in which both, DCDM and DR, have reached thermodynamical equilibrium
    \begin{equation}
        \omega_{\mathrm{mix}} = \frac{\rho_{\mathrm{DR}}}{3(\rho_{\mathrm{dcdm}}+\rho_{\mathrm{DR}})}.
    \end{equation}
    
\subsubsection{Temperature of different species}

\begin{figure}[htbp!]
        \centering
        \includegraphics[scale=0.5]{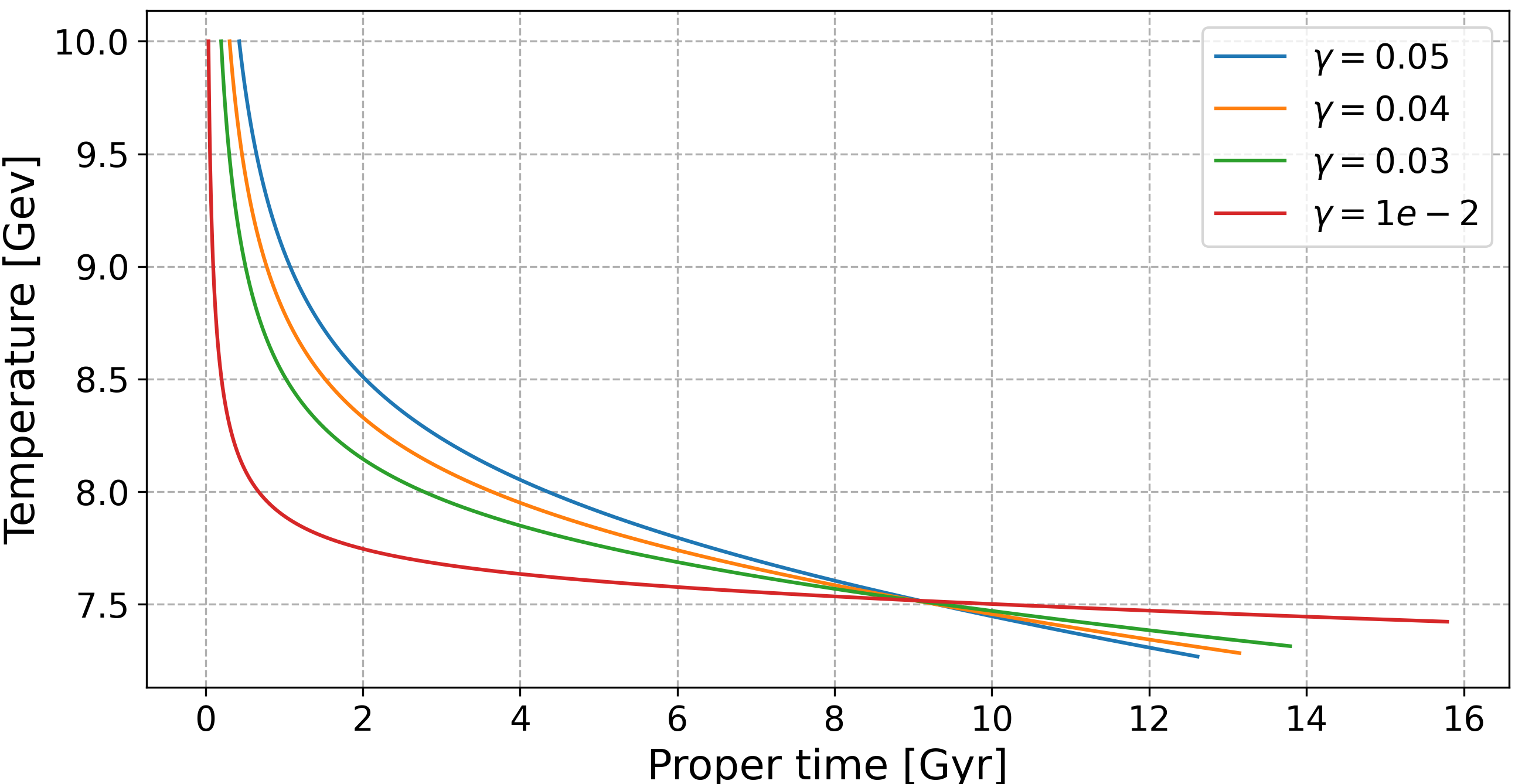}
        \includegraphics[scale=0.5]{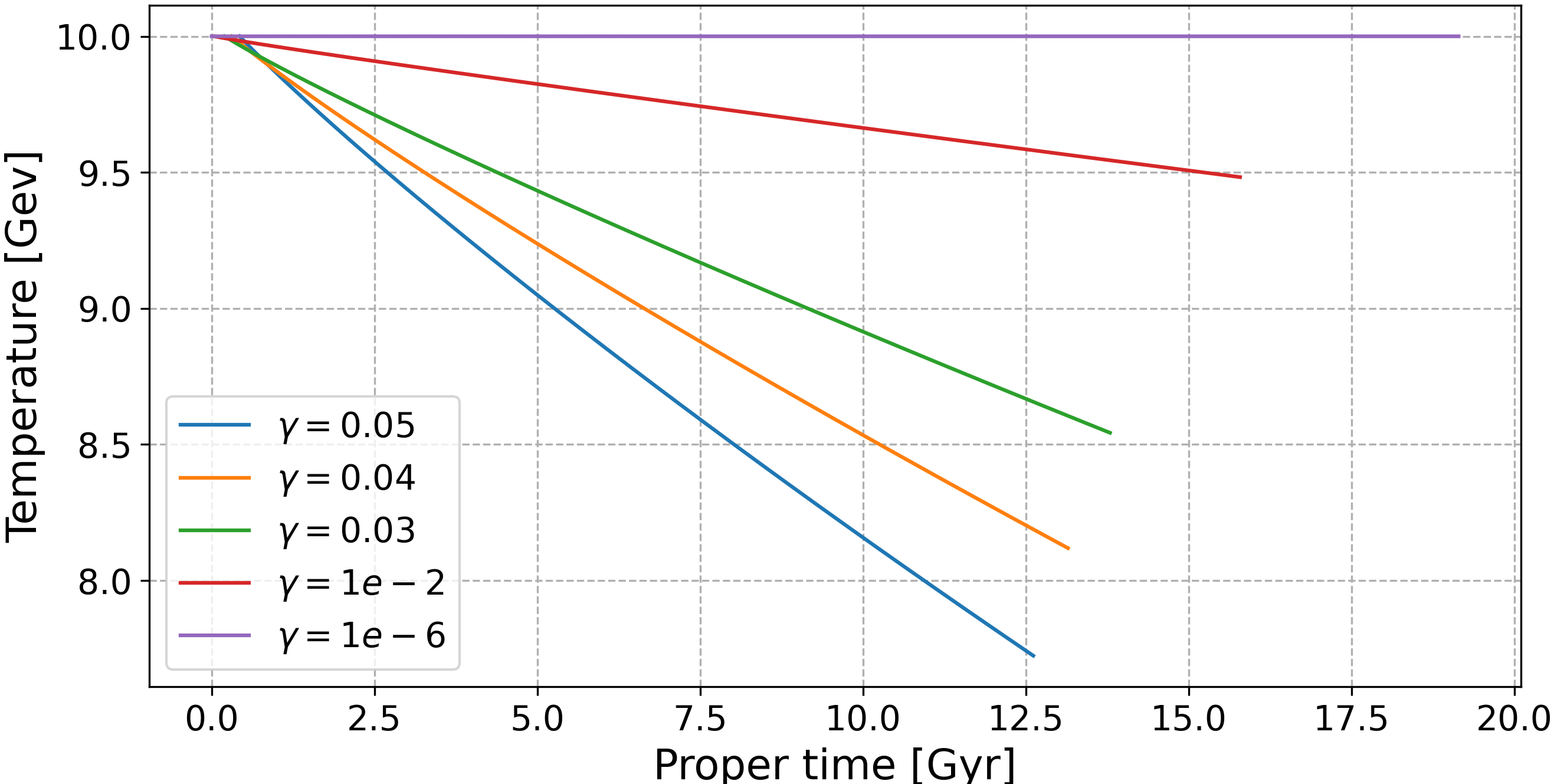}
        \includegraphics[scale=0.5]{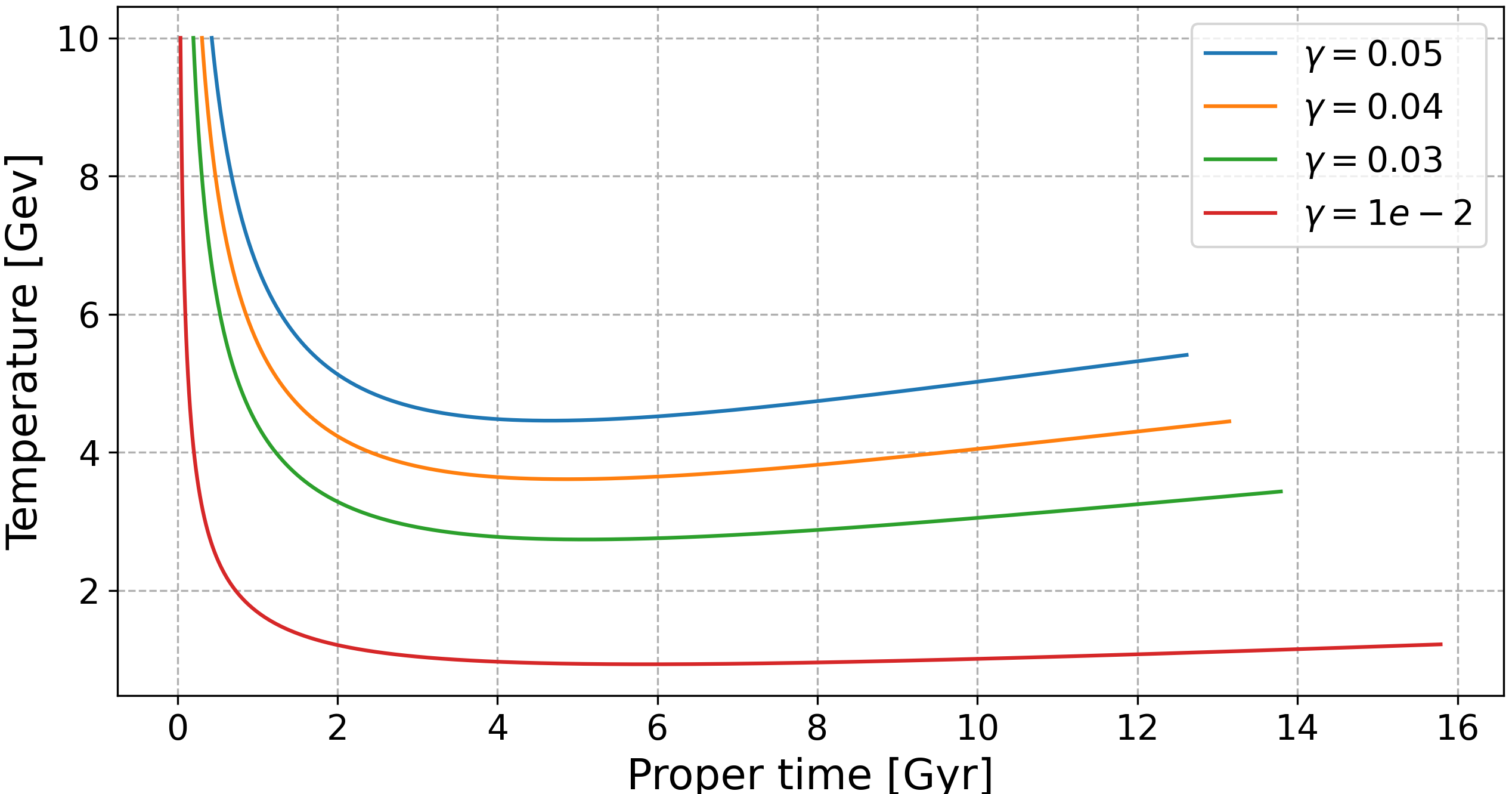}
        \caption{Temperature evolution. The upper panel exhibits the temperature of the mixture of fluids, the middle panel corresponds to the DCDM sector and the lower panel shows the temperature behavior of the DR sector.}
        \label{fig:temperature}
\end{figure}

Once the equation of state parameter for each species is obtained, and defining $\mathcal{H}\equiv a'/a$, where the prime denotes derivatives with respect to conformal time, from equation \ref{tempe} we can write the evolution equation for the temperature of DCDM
\begin{equation}
        \frac{T'_{\mathrm{dcdm}}}{T_{\mathrm{dcdm}}} = \frac{-\left(\frac{3\mathcal{H}}{a}-\epsilon_{\mathrm{dcdm}}\right)}{\frac{1}{a}+\frac{\epsilon\rho_{\mathrm{dcdm}}(2+\omega_{\mathrm{dcdm}})}{\rho'_{\mathrm{dcdm}}}}\omega_{\mathrm{dcdm}},
\end{equation}
and for DR we obtain the following expression
      \begin{equation}
        \frac{T'_{\mathrm{DR}}}{T_{\mathrm{DR}}} = \frac{\left( \Gamma_{\mathrm{DR}}-\frac{3\mathcal{H}}{a}\right)\omega_{\mathrm{DR}}}{\frac{1}{a}-\frac{\epsilon_{\mathrm{dcdm}}-\Gamma_{\mathrm{DR}}\rho_{\mathrm{DR}}(1+\omega_{\mathrm{DR}})}{\rho'_{\mathrm{DR}}}}
    \end{equation}
Due to the fact that for the mixed fluid, the total energy is conserved, implying conservation of the particle number, we must have $\Gamma_{\mathrm{mix}} = 0$, leading to the usual equation for the evolution of the temperature 
     \begin{equation}
        \frac{T'_{\mathrm{mix}}}{T_{\mathrm{mix}}} = -3\mathcal{H}\omega_{\mathrm{mix}},
    \end{equation}
which corresponds to the equilibrium case. Figure \ref{fig:temperature} shows the numerical solutions to the preceding differential equations. Initial conditions for the temperature were set around the freeze-out value of the annihilation of dark matter in order to get a thermal relic density consistent with estimates reported by the Planck 2018 collaboration \textbf{\citep{Planck2018}}. Specifically,  we set $T_i$ values within a range $(M/25,M/10)$, where $M\simeq 100 GeV$ corresponds to the dark matter candidate mass. It was verified that the resulting solutions are insensitive to the choice of $T_i$ laying within such range. As can be seen, due to the particle production, the evolution of the temperature differs from the standard case, where $T_{\mathrm{cdm, r}}=T_{0}$, being $T_{0}$ a positive constant. In this scenario, the temperature of the cosmic fluids decreases as the universe expands. However, this behavior for the temperature is typical of scenarios in which the interchange of energy is allowed, see for instance Refs. \textbf{\citep{grandon, Wang_2016}}. Notice that for the mixture of fluids, the variations of temperature are small for times of the order of the age of the universe.   

\subsubsection{The entropy}

\begin{figure}[htbp!]
        \centering
        \includegraphics[scale=0.5]{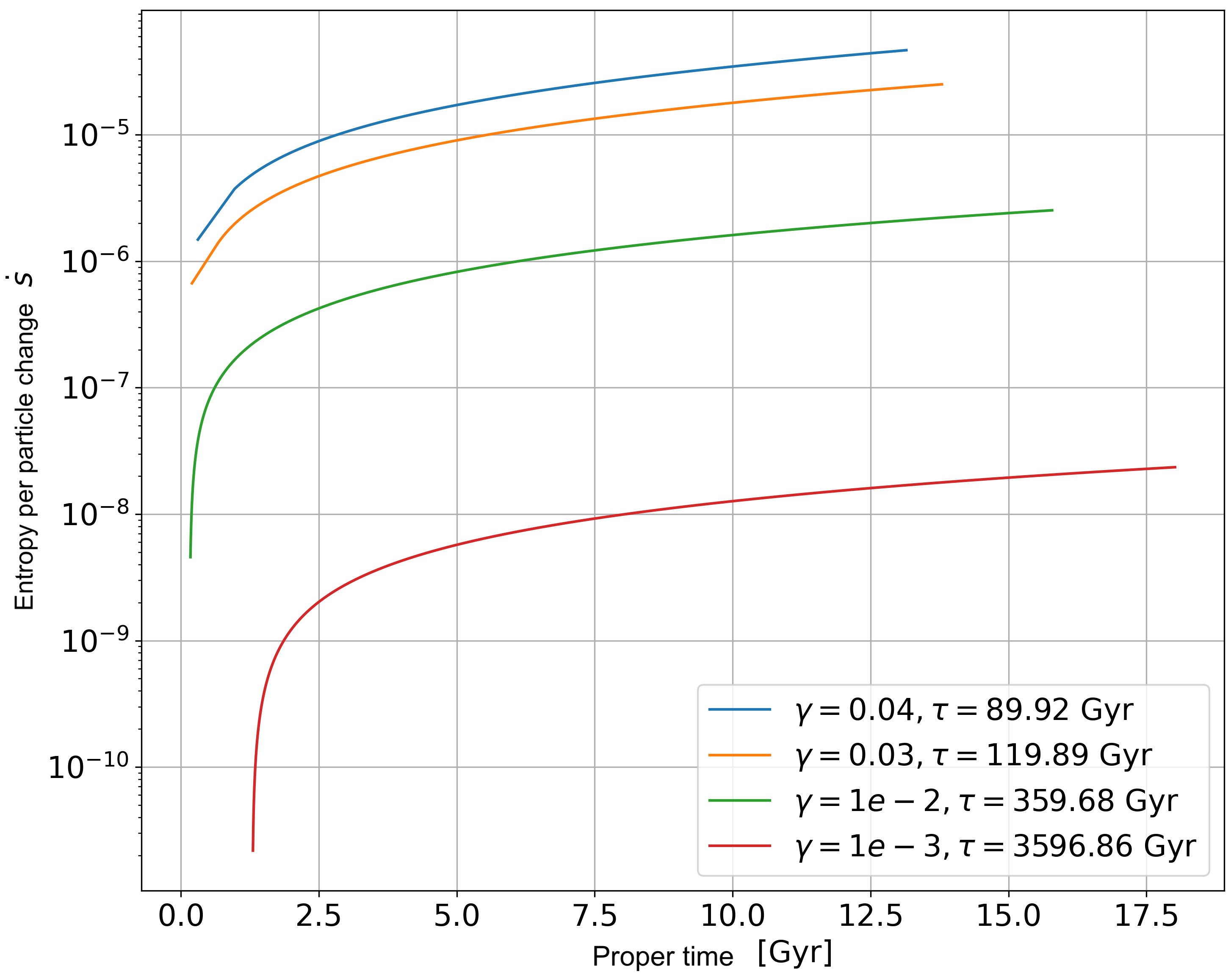}
        \caption{Proper time derivative of entropy per particle for different $\gamma$ (and DCDM lifetime) values.}
         \label{fig:entropy}

    \end{figure}

Independently of the cosmological model, the second law of thermodynamics holds at cosmic scales for a homogeneous and isotropic universe \textbf{\citep{manuel}}. In this sense, we verify the validity of the second law of thermodynamics in our cosmological model. We proceed by calculating the proper time derivative of entropy per particle for each species and plot the total production of entropy. From the first law of thermodynamics \textbf{\citep{8157526, maartens}}
\begin{equation}
    T_{A}ds_{A} =d\left(\frac{\rho_{A}}{n_{A}}\right) +p_{A}d\left(\frac{1}{n_{A}}\right),
\end{equation}
the contribution coming from the DCDM and DR sectors can be written as 
\begin{equation}
    \dot{s} = \frac{\rho_{\mathrm{dcdm}}\epsilon_{\mathrm{dcdm}}\omega_{\mathrm{dcdm}}}{n_{\mathrm{dcdm}}T_{\mathrm{dcdm}}} + \frac{\rho_{\mathrm{DR}}(\epsilon_{\mathrm{dcdm}} - \Gamma_{\mathrm{DR}}(1+\omega_{\mathrm{DR}}))}{n_{\mathrm{DR}}T_{\mathrm{DR}}},\label{entropia_ambas}
\end{equation}
which depends on the decay rates $\Gamma$ and $\epsilon$, as expected. In figure \ref{fig:entropy} we can see that the entropy per particle is an increasing function of proper time, verifying the second law of thermodynamics for some different values of the parameter $\gamma$. As can be seen, $\dot{s} \geq 0$, implying that the cosmic evolution in this scenario corresponds to an irreversible process. It is worth noting that entropy production at cosmic scales plays a significant role, as it contributes to the accelerated expansion of the universe. This phenomenon provides a more robust cosmological model for understanding the evolution of our universe, without worsening the already known tensions in cosmology, as demonstrated in the works referenced in \textbf{\citep{entro1, entro2}}. Therefore, the set of solutions obtained in this work leads to a consistent thermodynamic cosmological scenario.    

\section{Conclusions}
\label{sec:final}

This study examines the thermodynamic properties that emerge from a novel set of solutions within a cosmological framework where dark matter undergoes decay into dark radiation. Some comments are in order:

The decay process introduces an irreversible component to the cosmic evolution, leading to a continuous increase in the total entropy of the dark sector as the universe expands. This entropy production is directly proportional to the particles' decay rates, as found in other cosmological models in which particle production is allowed, see for instance Ref. \textbf{\citep{waga}}.

We have demonstrated that the standard adiabatic expansion of the universe is modified by the energy transfer between the dark matter and dark radiation components. The effective equation of state of the combined dark sector exhibits a time-dependent behavior. Although the dark radiation sector may exhibit a negative equation of state parameter, the model necessitates an additional component to achieve accelerated expansion. Further work is needed to fully explore the implications of the analytical solutions obtained in this model at early times, particularly during the epoch of matter-radiation equality, and to investigate possible connections with other open questions in cosmology, such as the $H_{0}$ tension and the $\sigma 8$ problem since this kind of cosmological scenario mitigates these issues, as found in Ref. \textbf{\citep{diff2}}. We will report this elsewhere. 

The thermodynamic analysis reveals that this model satisfies the second law of thermodynamics. An interesting case could be provided by considering a generalized second law of thermodynamics in which the inclusion of the apparent horizon entropy must be taken into account, see for instance Ref. \textbf{\citep{us}}, wherein matter creation phenomena confined within the apparent horizon generate a quintessence-phantom cosmological paradigm. We leave this subject open for future investigation.

The derivation of novel solutions and verification of their physical consistency across diverse cosmological scenarios bears significant implications for both theoretical cosmology frameworks and observational constraints on dark matter properties. Future observations of large-scale structure and precise measurements of the cosmic microwave background could potentially constrain the decay parameter and provide insights into the nature of dark matter and its interactions, as found in Ref. \textbf{\citep{dmdm2}}.

\acknowledgments
JJJ acknowledges SECIHTI-M\'exico for the master studies scholarship program. AAAL and MC work has been partially supported by S.N.I.I. (SECIHTI-M\'exico). JJJ and AAAL acknowledge VIEP-BUAP for supporting this work by means of the "Proyectos VIEP 2024" program with I.D. 00497.

\bibliographystyle{ieeetr}
\bibliography{biblio}
\end{document}